\documentstyle[aps,epsfig,floats]{revtex}
\def\lsi{\raise0.3ex\hbox{$<$\kern-0.75em\raise-1.1ex\hbox{$\sim$}}}
\def\gsi{\raise0.3ex\hbox{$>$\kern-0.75em\raise-1.1ex\hbox{$\sim$}}}
\newcommand{\lsim}{\mathop{\lsi}}

\begin{document}
\draft
\twocolumn[\hsize\textwidth\columnwidth\hsize\csname
@twocolumnfalse\endcsname

\title{Grand Unification signal from ultrahigh energy cosmic rays?}
\author{
Z.~Fodor and S.D.~Katz}
\address{Institute for Theoretical Physics, E\"otv\"os University, P\'azm\'any 1, H-1117 Budapest, Hungary}

\date{\today}
\maketitle

\vspace*{-4.0cm}
\noindent
\hfill \mbox{ITP-Budapest 560}
\vspace*{3.8cm}

\begin{abstract} \noindent
The spectrum of ultrahigh energy (above $\approx 10^{9}$~GeV)
cosmic rays is consistent with the decay of 
GUT scale particles. The
predicted mass is $m_X=10^b$~GeV, where $b=14.6_{-1.7}^{+1.6}$.
\end{abstract}

\vspace*{0.2cm}
\pacs{PACS numbers: 98.70.Sa, 96.40.Pq, 13.87.Fh}
\vskip1.3pc]

The interaction of protons with photons of the cosmic microwave 
background radiation (CMBR) predicts a sharp drop in the cosmic ray flux above
the Greisen-Zatsepin-Kuzmin (GZK) cutoff around $5\cdot 10^{19}$~eV \cite{GZK66}. The 
available data shows no such drop. About 20 events above $10^{20}$~eV
were observed by experiments such as Akeno Giant Air Shower Array (AGASA) 
\cite{AGASA}, Fly's Eye \cite{FLY}, Haverah Park \cite{HAVERAH},
Yakutsk \cite{YAKUTSK} and HiRes \cite{HIRES}. 
Future experiments, particularly Pierre Auger \cite{PAUG}, will 
have a much higher statistics.

Usually it is assumed that 
at these energies the galactic and extragalactic (EG) magnetic 
fields do not affect the orbit of the cosmic rays, thus they 
should point back to their origin within a few degrees. 
Though there are clustered events \cite{Hea96,Uea00}
the overall distribution is practically isotropic
\cite{BM99},
which usually ought to be interpreted as a signature for EG 
origin. 

Since above
the GZK energy the attenuation length of particles is a few tens
of megaparsecs \cite{YT93,BS00,AGNM99,SEMPR00} 
if an ultrahigh energy cosmic ray (UHECR) is
observed on Earth it must be produced in our vicinity (except for UHECR
scenarios based on weakly interacting particles, e.g. neutrinos 
\cite{DKDM00}).
Sources of EG origin (e.g. AGN \cite{BS87}, topological
defects \cite{HSW87} or the local supercluster \cite{BG79}) should 
result in a GZK cutoff, which is in disagreement with experiments.  
It is generally believed \cite{B99} that there is no 
conventional astrophysical explanation for the observed UHECR spectrum.

An interesting idea suggested by refs.\cite{BKV97,KR98} is that 
superheavy particles (SP) as dark matter could be the source of UHECRs. 
(Note, that metastable relic SPs were proposed 
\cite{ELN90} before the observation of UHECRs beyond 
the GZK cutoff.) In 
\cite{KR98} EG SPs were studied.
Ref. \cite{BKV97} made a crucial observation and analyzed the
decay of SPs concentrated in the 
halo of our galaxy. They used  the modified leading logarithmic
approximation (MLLA) \cite{MLLA} for ordinary QCD and for supersymmetric QCD
\cite{BK98}. 
A good agreement of the EG spectrum with observations was noticed in \cite{BBV}.
Supersymmetric QCD is treated as the strong regime of the
minimal supersymmetric standard model (MSSM).
To describe the decay spectrum more 
accurately HERWIG Monte-Carlo was used in QCD \cite{BS98} and discussed
in supersymmetric QCD \cite{S00,Rubin}, resulting in 
$m_X \approx 10^{12}$ GeV and $\approx 10^{13}$ GeV for the SP mass in  
SM and in MSSM, respectively. 

SPs are very 
efficiently produced by the various mechanisms at post inflatory 
epochs \cite{B00}. Note, that our analysis of SP decay 
covers a much broader class of possible sources. 
Several non-conventional UHECR sources
(e.g. EG long ordinary strings \cite{VAH98} or galactic
vortons \cite{MS98}, monopole-antimonopole pairs connected by strings
\cite{BO99}) produce the same UHECR spectra as decaying SPs. 

In this letter we study the scenario that the UHECRs are coming
from decaying SPs and we determine the mass of this $X$ 
particle $m_X$ by a detailed analysis of the observed UHECR spectrum. 
We discuss both possibilities that the UHECR protons are 
produced in the halo of our galaxy and that they are of EG
origin and their propagation is affected by CMBR. Here we do not investigate
how can they be of halo or EG origin, we just analyze their effect on the observed
spectrum instead. We assume that the SP decays into two quarks (other decay modes 
would increase $m_X$ in our conclusion). 
After hadronization these quarks yield protons. The result is characterized 
by the fragmentation function (FF) $D(x,Q^2)$ which gives the number of produced 
protons with momentum fraction $x$ at energy scale $Q$.
For the proton's FF at present accelerator
energies we use ref. \cite{BKK95}. We evolve
the FFs in ordinary \cite{DGLAP} and 
in supersymmetric \cite{JL83} QCD to the energies of the 
SPs. This result can be combined with the
prediction of the MLLA technique 
, which gives 
the initial spectrum of UHECRs at the energy $m_X$. 
Altogether we study four different models:
halo-SM, halo-MSSM, EG-SM and EG-MSSM. 


Ref. \cite{DMS98} showed that both AGASA and Fly's
Eye data demonstrated a change of composition, a shift from heavy
--iron-- at $10^{17}$ eV to light --proton-- at  $10^{19}$ eV. Thus
the UHECRs are most likely
to be dominated by protons and in our analysis we use them exclusively.

The FF of the proton can be determined from 
present experiments \cite{BKK95}. (Note, that QCD event generators e.g.
HERWIG \cite{HERWIG} predict the overall proton multiplicity 
correctly, however they describe the large $x$ region of the 
FF inaccurately.)
The FFs 
at $Q_0$ energy scale are
$D_i(x,Q_0^2)$,
where $i$ represents the different partons (quark/squark or gluon/gluino).
The FFs can not be determined
in perturbative QCD; however, their evolution in $Q^2$ 
is governed by the DGLAP equations \cite{DGLAP}:
\begin{equation} 
{\partial D_i(x,Q^2) \over \partial \ln Q^2}= 
\frac{\alpha_s(Q^2)}{2\pi}
\sum_j \int_x^1 {dz \over z} P_{ji}(z,\alpha_s(Q^2))D_j(\frac{x}{z},Q^2),
\end{equation} 
One can interpret
$P_{ji}(z)$, the splitting function, as the probability density that
a parton $i$ produces a parton $j$ 
with momentum fraction $z$.
Analogous evolution equations can be obtained by using coherent branching
(angular ordering) for
the emitted gluons \cite{Webber,Rubin}. We use this
technique too. Results of direct Monte-Carlo jet simulations are also available
(cf. \cite{BKjet}). We include the uncertainties coming from the different
choices of evolution in our final error estimates.
We solve the DGLAP equations numerically with the 
conventional QCD (SM case) splitting functions and with the supersymmetric 
(MSSM case) ones \cite{JL83}. For the top and the MSSM partons
we used the FFs of ref. \cite{Rubin}. While solving the DGLAP equations each 
parton is included at its own threshold energy.
Table \ref{frag_tab} shows all the FFs we used.

\begin{table}[t]
\begin{center}\begin{tabular}{l|l|l|l|l}
flavor&Q(GeV)&$N$&$\alpha$&$\beta$\\
\hline
$u=2d$ &1.41&0.402&-0.860&2.80\\
\hline
$s$    &1.41&4.08&-0.0974&4.99\\
\hline
$c$    &2.9&0.111&-1.54&2.21\\
\hline
$b$    &9.46&40.1&0.742&12.4\\
\hline
$t$    &350&1.11&-2.05&11.4\\
\hline
$g$    &1.41&0.740&-0.770&7.69\\
\hline
$\tilde{q}_i, \tilde{g}$&1000&0.82&-2.15&10.8\\
\end{tabular}
\vspace{0.3cm}
\caption{\label{frag_tab}
{The fragmentation functions of the different partons using the 
parametrization $D(x)=Nx^\alpha (1-x)^\beta$ at different energy scales
(second column).
}}
\end{center}\end{table}

\begin{figure}\begin{center}
\epsfig{file=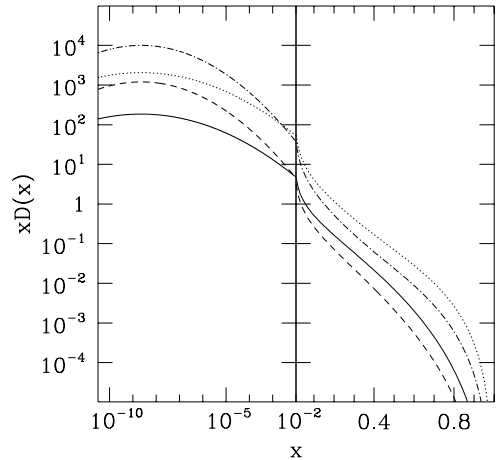,width=6.8cm}
\caption{\label{fragmentation}
{The FFs averaged over the quark flavors
at $Q=10^{16}$ GeV for proton/pion in SM (solid/dotted line)  and in MSSM 
(dashed/dashed-dotted line)
in the relevant $x$ region. To show both the small and large $x$
behavior we change from logarithmic scale to linear at $x=0.01$.
}}
\end{center}\end{figure}

At small values of $x$, multiple soft gluon emission can be 
described by the MLLA \cite{MLLA}.
This gives the shape of the total hadronic FF  for soft particles (not distinguishing 
individual hadronic species)
$xF(x,Q^2)\propto \exp \left[ -\ln(x/x_m)^2/(2\sigma^2)\right],$
which is peaked at $x_m = \sqrt{\Lambda/Q}$ with 
$2\sigma^2= A\ln^{3/2}(Q/\Lambda)$. According to \cite{BK98}
the values of $A$ are 
$\sqrt{7/3}/6$ and $1/6$ for SM and MSSM, respectively.
The MLLA describes the observed hadroproduction quite accurately
in the small $x$ region \cite{DELPHI}.
For large values of $x$ the MLLA should not be
used. We smoothly connect the solution for the FF
obtained by the DGLAP equations 
and the MLLA result at a given $x_c$ value.
Our final result on $m_X$
is rather insensitive to the choice of $x_c$, the uncertainty is included
in our error estimate. We also determined the FF of the pion.
Fig. \ref{fragmentation} shows the FF for the proton and 
pion at $Q=10^{16}$~GeV in  SM and MSSM.


UHECR protons produced in the halo of our galaxy can propagate
practically unaffected and the production spectrum should be
compared with the observations. 

Particles of EG origin
and energies above $\approx 5\cdot 10^{19}$ eV loose a large fraction 
of their energies due to interactions with CMBR 
\cite{GZK66}. This effect can be quantitatively described by the
function $P(r,E,E_c)$, the probability
that a proton created at a distance $r$ with energy $E$ arrives
at Earth above the threshold energy $E_c$ \cite{BW99}.
This function has been calculated for a wide range
of parameters in \cite{FK00}, which we use in the present calculation.
The original UHECR spectrum 
is changed at least by two different ways: (a) there should be a 
steepening due to the GZK effect; (b) particles loosing their
energy are accumulated just before the cutoff and produce a bump. 
We study the observed spectrum by
assuming a uniform source distribution for UHECRs.

\begin{figure}[t]\begin{center}
\epsfig{file=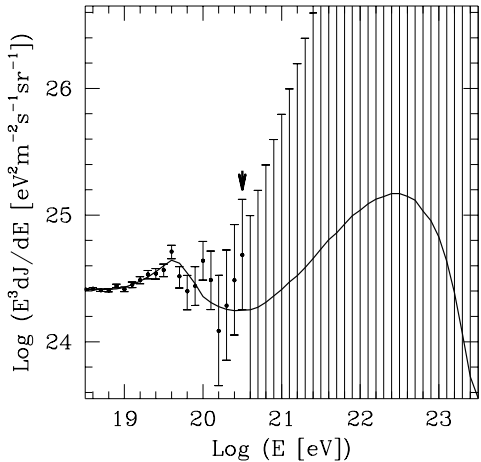,width=6.8cm}
\caption{\label{spect}
{The available UHECR data with their error bars
and the best fit from a decaying SP.
Note that there are no events above $3 \times 10^{20}$~eV 
(shown by an arrow). 
Nevertheless the experiments are sensitive even in this region. Zero event
does not mean zero flux, but a well defined upper bound for the flux 
(given by the Poisson distribution).
Therefore the experimental value of the
integrated flux is in the ''hatched'' region with 68\% confidence level. 
(''hatching'' is a set of individual 
error bars; though most of them are too large to be depicted in full) 
Clearly, the error bars are large enough to be consistent with the SP decay.
}}
\end{center}\end{figure}

Our analysis includes the published and the unpublished
(from the www pages of the experiments on 17/08/00) UHECR data 
of \cite{AGASA,FLY,HAVERAH,HIRES}. Due to normalization difficulties
we did not use the Yakutsk \cite{YAKUTSK} results. 
We also performed the analysis using the AGASA data only and found the 
same value (well within the error bars) for $m_X$.
Since the decay of SPs results 
in  a non-negligible flux for lower energies $\log (E_{min}/\mbox{eV})=18.5$ 
is used as a lower end for the UHECR spectrum. Our results are
insensitive to the definition of the upper end (the flux is
extremely small there) for which we choose $\log (E_{max}/\mbox{eV})=26$.
As it is usual we divided each logarithmic unit into ten bins. The
integrated flux
gives the total number of events in a bin. The uncertainties of the 
measured energies are about 30\% which is one bin. Using a Monte-Carlo method
we included this uncertainty in the final error estimates.
The predicted number of events in a bin is given by
\begin{equation}\label{flux}
N(i)=\int_{E_i}^{E^{i+1}}\left[A \cdot E^{-3.16}+B\cdot j(E,m_X)\right],
\end{equation}
where $E_i$ is the lower bound of the i$^{th}$ energy bin. The first 
term describes the data below $10^{19}$~eV according to
\cite{AGASA}, where the SP decay gives negligible contribution.
The second one corresponds to the spectrum of the decaying
SPs.
A and B are normalization factors.

The expectation value for the number of events in a bin is given
by eqn. (\ref{flux}) and it is Poisson distributed. To
determine the most probable $m_X$ value we used the maximum-likelihood
method by minimalizing the $\chi^2(A,B,m_X)$ for
Poisson distributed data \cite{PDG}
\begin{equation} \label{chi}
\chi^2=\sum_{i=18.5}^{26.0}
2\left[ N(i)-N_o(i)+N_o(i)\ln\left( N_o(i)/N(i)\right) \right],
\end{equation}
where $N_o(i)$ is the total number of observed events in the i$^{th}$ 
bin. In our fitting procedure we have three parameters: $A,B$ and $m_X$.
The minimum of the $\chi^2(A,B,m_X)$ function is $\chi^2_{min}$ 
at $m_{X min}$ which is 
the most probable value for the mass, whereas
$\chi^2(A',B',m_X)\equiv \chi^2_o(m_X)=\chi^2_{min}+1$
gives the one-sigma (68\%) confidence interval for $m_X$. 
Here $A',B'$ are defined in 
such a way that the $\chi^2(A,B,m_X)$ function is minimalized in $A$ 
and $B$ at fixed $m_X$.
Fig. \ref{spect} shows the measured UHECR spectrum and the best fit in the
EG-MSSM scenario.
The first bump of the fit represents particles produced at
high energies and accumulated just above the GZK cutoff due to their energy
losses. The bump at higher energy is a remnant of $m_X$. In the halo
models there is no GZK bump, so the relatively large $x$ part of the FF moves
to the bump around $5\times 10^{19}$~GeV resulting in a much smaller $m_X$ than
in the EG case. An interesting feature of the GZK effect is that the shape of
the produced GZK bump is rather insensitive to the injected spectrum so the
dependence of $\chi^2$ on the choice of the FF is small.
The experimental data is far more accurately described by the 
GZK effect (dominant feature of the EG fit) than by the FF itself (dominant for 
halo scenarios).

   

\begin{figure}\begin{center}
\epsfig{file=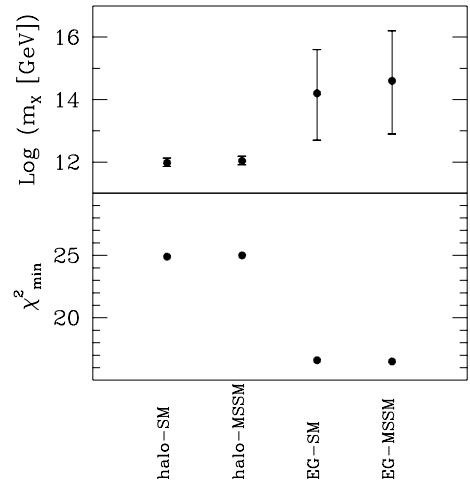,width=6.8cm}
\caption{\label{result}
{The most probable values for the mass of the decaying
ultra heavy dark matter with their error bars and the
total $\chi^2$ values. Note that 21 bins contain nonzero number of events
and eqn.(\ref{flux}) has 3 free parameters.
}}
\end{center}\end{figure}

To determine the most probable value for the mass of the 
SP we studied 4 scenarios. Fig. \ref{result} contains
the $\chi^2_{min}$ values and the most
probable masses with their errors for these scenarios.
(The uncertainties coming from 
the FFs are included in our error estimates on $m_X$.)

The UHECR data favors the EG-MSSM scenario. The
goodnesses of the fits for the halo models are far worse. 
The SM and MSSM cases do not differ significantly. 
The most important message is that the masses of the best fits 
(EG cases)
are compatible within the
error bars with the MSSM gauge coupling unification GUT scale \cite{ABF92}.

The SP decay will also produce a huge number of pions which will decay into 
photons. Our spectrum contains 94\% of pions and 6\% of protons.
This $\pi /p$ ratio is in agreement with
\cite{Sigl99,BS00} which showed that
for different classes of models $m_X \lsim 10^{16}$~GeV , which is the upper boundary
of our confidence intervals, the generated
gamma spectrum is still consistent with the observational constraints. 
We performed the whole analysis including the pion produced $\gamma$-s in eqn.
(\ref{chi}). The results agree with our results of Fig. \ref{result} within 
errorbars, which is easy to understand. For the EG case high energy $\gamma$-s
dominate at energies where the observed flux is zero \cite{BBV}. For the halo
case the agreement is resulted by the similarity (except normalization) 
between $D_p$ and $D_\pi$ (cf. Fig. \ref{fragmentation}).

In the near future the UHECR statistics will probably be increased by an
order of magnitude \cite{PAUG}. Performing our analysis for such a
statistics the uncertainty of $m_X$ was found to
be reduced by two orders of magnitude.

Since the decay time should be at least
the age of the universe it might happen that such SPs
overclose the universe. Due to the large mass of the SPs a single decay 
results in a large number of UHECRs, thus a relatively small 
number of SPs can describe the observations. We checked that in 
all of the four scenarios the minimum density required for the 
best-fit spectrum is more 
than ten orders of magnitude smaller than the critical one.

Details will be presented in a subsequent paper
\cite{FK01}.

We thank B.A. Kniehl for providing us with the proton's FF prior
to its publication and F. Csikor for useful comments.
This work was partially supported by Hungarian Science Foundation
grants No. OTKA-T29803/T22929-FKP-0128/1997.


\begin{thebibliography}{99}
\bibitem{GZK66} K.Greisen, Phys.Rev.Lett. {\bf 16}, 748 (1966); 
G.T.Zatsepin, V.A.Kuzmin, Pisma Zh.Exp.Teor.Fiz. {\bf 4}, 114 (1966).
\bibitem{AGASA} M. Takeda et al., Phys. Rev. Lett. {\bf 81}, 1163 (1998);
astro-ph/9902239; www-akeno.icrr.u-tokyo.ac.jp/AGASA/
\bibitem{FLY} D.J. Bird et al., Phys. Rev. Lett. {\bf 71}, 3401 (1993);
Astrophys J. {\bf 424}, 491 (1994); ibid {\bf 441}, 144 (1995).
\bibitem{HAVERAH} M.A. Lawrence, R.J.O. Reid and A.A. Watson,
J. Phys. {\bf G17}, 773 (1991).
\bibitem{YAKUTSK} N.N.~Efimov et al., "Proc. Astrophysical Aspectes \dots",
M.~Nagano and F.~Takahara, World Sci., Singapore, 1991.
\bibitem{HIRES} D. Kieda et al., to appear in Proc. of the 26th ICRC, Salt Lake,
1999; www.physics.utah.edu/Resrch.html
\bibitem{PAUG} M. Boratav, Nucl. Phys. Proc. {\bf 48}, 488 (1996);
C.K. Guerard, ibid {\bf 75A}, 380 (1999);
X. Bertou, M. Boratav, A. Letessier-Selvon, astro-ph/0001516.
\bibitem{Hea96} N. Hayashida et al., Phys. Rev. Lett. {\bf 77}, 1000 (1996).
\bibitem{Uea00} Y. Uchihori et al., Astropart. Phys. {\bf 13}, 151 (2000).
\bibitem{BM99} S.L. Dubovski, P.G. Tinyakov, JETP Lett. {\bf 68}, 107 (1998); 
V.Berezinsky, A.A. Mikhailov, Phys. Lett. {\bf B449}, 61 (1999); 
C.A. Medina Tanco, A.A. Watson, Astrop. Phys. {\bf 12}, 25 (1999).
\bibitem{YT93} S. Yoshida, M. Teshima, Prog. Theor. Phys. 
{\bf 89}, 833 (1993);
F.A. Aharonian, J.W. Cronin, Phys. Rev. {\bf D50}, 1892 (1994);
R.J. Protheroe, P. Johnson, Astropart. Phys. {\bf 4}, 253 (1996).
\bibitem{BS00} P. Bhattacharjee and G. Sigl, Phys. Rep. {\bf 327}, 109 (2000).
\bibitem{AGNM99} A. Achterberg et al., astro-ph/9907060.
\bibitem{SEMPR00} T. Stanev et al., astro-ph/0003484.
\bibitem{DKDM00} G. Domokos, S. Nussinov, Phys. Lett. {\bf B187}, 372 (1987);
D. Fargion, B. Mele, A. Salis, Astrophys. J. {\bf 517}, 725 (1999);
T.J. Weiler, Astropart. Phys. {\bf 11} (1999) 303, Astropart. Phys. 
{\bf 12}, 379 (2000) (Erratum); 
G.~Domokos, S.~Kovesi-Domokos,
Phys.\ Rev.\ Lett.\ {\bf 82}, 1366 (1999).
\bibitem{BS87} P.L. Biermann, P.A. Strittmatter, Astrophys. J. 
{\bf 322}, 643 (1987).
\bibitem{HSW87} C.T. Hill, D.N. Schramm, T.P. Walker, Phys. Rev. 
{\bf D36}, 1007 (1987); 
P. Bhatacharjee, C.T. Hill, D.N. Schramm, Phys. Rev. Lett.
{\bf 69}, 56 (1992); G. Sigl astro-ph/9611190; V. Berezinsky, A. Vilenkin,
astro-ph/9704257.
\bibitem{BG79} V.S. Berezinsky, S.I. Grigorieva, in Proc. of the 16th
Int. Cosmic Ray  Conf., Kyoto 1979, Vol. {\bf 2.} p. 81.
\bibitem{B99} R.D. Blandford, Phys. Scr. {\bf T85}, 191 (2000). 
\bibitem{BKV97} V. Berezinsky, M. Kachelrie{\ss}   and A. Vilenkin, 
Phys. Rev. Lett. {\bf 79}, 4302 (1997);
\bibitem{KR98} V.A. Kuzmin, V.A. Rubakov, Phys. Atom. Nucl. 
{\bf 61}, 1028 (1998).
\bibitem{ELN90} J. Ellis et al., Phys. Lett.
{\bf B247}, 257 (1990);
Nucl. Phys. {\bf B373}, 399 (1992);
P. Gondolo, G.B. Gelmini, S. Sarkar, Nucl. Phys. {\bf B392}, 111 (1993).
\bibitem{MLLA} Ya.I. Azimov, Yu.L. Dokshitzer, V.A. Khoze, 
S.I. Troyan, Phys. Lett. 
{\bf B165}, 147 (1985); Z. Phys. {\bf C27}, 65 (1985);
$ibid$ {\bf C31}, 213 (1986); C.P. Fong, B.R. Webber, Nucl. Phys. 
{\bf B355}, 54 (1991).
\bibitem{BK98} V.Berezinsky, M.Kachelrie{\ss}, Phys.Lett.
{\bf B434}, 61 (1998).
\bibitem{BBV} V.~Berezinsky, P.~Blasi, A.~Vilenkin, Phys. Rev. {\bf D58},
103515 (1998).
\bibitem{BS98} M. Birkel and S. Sarkar, Astropart. Phys. {\bf 9}, 297 (1998).
\bibitem{S00} S. Sarkar, hep-ph/0005256.
\bibitem{Rubin}  N. Rubin, www.stanford.edu/$^\sim$nrubin/Thesis.ps
\bibitem{B00} for a review see V. Berezinsky, astro-ph/0001163.  
\bibitem{VAH98} G. Vincent, N. Antunes, M. Hindmarsh, Phys. Rev. Lett.
{\bf 80}, 2277 (1998); M. Hindmarsh hep-ph/9806469.
\bibitem{MS98} L. Masperi, G. Silva, Astrop. Phys. {\bf 8}, 173 (1998).
\bibitem{BO99} J.J. Blanco-Pillado, K.D. Olum, astro-ph/9909143.
\bibitem{BKK95} J.Binnenwies, B.A.Kniehl, G. Kramer,
Phys. Rev. {\bf D52}, 4947 (1995);
B.A. Kniehl, G. Kramer, B. Potter, Phys. Rev. Lett. {\bf 85}, 5288 (2000);Nucl. Phys. {\bf B582}, 514 (2000);
\bibitem{DGLAP} V.N. Gribov, L.N. Lipatov, Sov. J. Nucl. Phys. 
{\bf 15}, 438 (1972); L.N. Lipatov, $ibid$ {\bf 20}, 94 (1975); 
G. Altarelli, G. Parisi, Nucl. Phys. {\bf B126}, 298 (1977); 
Yu.L. Dokshitzer, Sov. Phys. JETP {\bf 46}, 641 (1977).
\bibitem{JL83} S.K. Jones, C.H. Llewellyn Smith, 
Nucl. Phys. {\bf B217}, 145 (1983).
\bibitem{DMS98} B.R. Dawson, R. Meyhandan and K.M. Simpson, Astropart. 
Phys. {\bf 9}, 331 (1998).
\bibitem{HERWIG} G. Marchesini et al. Comp. Phys. Comm. {\bf 67}, 465 (1992). 
\bibitem{Webber}
G.~Marchesini and B.~R.~Webber, Nucl.\ Phys.\ {\bf B330}, 261 (1990).
S.~Catani, B.~R.~Webber and G.~Marchesini, Nucl.\ Phys.\ 
{\bf B349}, 635 (1991);
\bibitem{BKjet}
V.~Berezinsky and M.~Kachelriess, hep-ph/0009053.
\bibitem{DELPHI} see eg. P. Abreu et al. Phys. Lett. {\bf B459}, 397 (1999);
G. Abbiendi et al. hep-ex/0002012.
\bibitem{BW99} J.N. Bahcall and E. Waxman, hep-ph/9912326.
\bibitem{FK00} Z. Fodor, S.D. Katz, hep-ph/0007158.
\bibitem{PDG} C. Caso et al., Eur. Phys. J. {\bf C3}, 172 (1998).
\bibitem{ABF92} U. Amaldi, W. de Boer, H. Furstenau,
    Phys. Lett. {\bf B260}, 447 (1991).
\bibitem{Sigl99}
G.~Sigl, S.~Lee, P.~Bhattacharjee and S.~Yoshida, 
Phys. Rev. {\bf D59}, 043504 (1999).
\bibitem{FK01}
Z. Fodor, S.D. Katz, in preparation.

\end{thebibliography}
\end{document}